\documentclass[twocolumn,showpacs,preprintnumbers,amsmath]{revtex4}
\usepackage{amsfonts}
\usepackage{graphicx}
\usepackage{epsfig}
\usepackage{amsmath,amssymb}

\begin{document}

\title{Aspects of Diffeomorphism Invariant Theory of Extended Objects I: \\
The Relativistic Particle and its d-brane Cousins}
\author{V. G. Gueorguiev\footnote{On leave of absence from Institute of
Nuclear Research and Nuclear Energy, Bulgarian Academy of Sciences,
Sofia 1784, Bulgaria.}}
\affiliation{Lawrence Livermore National Laboratory, Livermore, CA 94550} 

\begin{abstract}
General structure of classical reparametrization-invariant matter systems,
mainly the relativistic particle and its $d$-brane generalization, are
studied. The exposition is in close analogy with the relativistic particle
in an electromagnetic field as reparametrization-invariant system. The
structure of a diffeomorphism invariant Lagrangian action for an extended
object ($d$-brane) embedded in a bulk space M is discussed. Our construction
uses first order homogeneous Lagrangians to achieve general covariance in
contrast to the constructions that use scalar Lagrangians along with metric
dependent integration measure. The framework contains intrinsically the
relativistic point particle, string theory, and Dirac-Nambu-Goto
Lagrangians.  In a natural way, the matter Lagrangian contains background interaction fields, such as a 1-form field, analogous to the electromagnetic vector potential,
and a metric tensor. The framework naturally suggests new classical interaction fields beyond electromagnetism and gravity. Construction of aninteraction field Lagrangian that is  background free and  consistent with the gauge symmetries presented
in the equations of motion for the matter is outlined.

\end{abstract}


% 03.50.-z Classical field theories
% 04.90.+e Other topics in general relativity and gravitation
% 11.10.-z Field theory
% 11.10.Ef Lagrangian and Hamiltonian approach
% 11.25.-w Strings and branes
% 11.25.Uv D branes
% 11.25.Wx String and brane phenomenology
% 11.27.+d Extended classical solutions;
% 11.90.+t Other topics in general theory of fields and particles
% 95.35.+d Dark matter

% 98.80.-k Cosmology
% 02.20.2a Group theory
% 02.40.2k Geometry, differential geometry, and topology
% 02.40.Yy Geometric mechanics
% 45.20.Jj in formalisms in classical mechanics
% 45.20.Jj Lagrangian and Hamiltonian mechanics
% 45.50.Ëj Dynamics and kinematics of a particle and a system of particles
% 12.90.+b Miscellaneous theoretical ideas and models 
% 04.60.Gw Covariant and sum-over-histories quantization

\maketitle

\section{Introduction}

Probing and understanding physical reality goes through a classical
interface that shapes our thoughts as classical causality chains. Therefore,
understanding the essential mathematical constructions in classical
mechanics and classical field theory is important, even though quantum
mechanics and quantum field theory are regarded as more fundamental than
their classical counterparts. Two approaches, the Hamiltonian and the
Lagrangian, are very useful in physics \cite{Kilmister 1967, Goldstain
1980,Gracia and Josep,Nikitin-string theory,Carinena 1995}. In general,
there is a transformation that relates these two approaches. For a
reparametrization-invariant theory, however, there are problems in changing
from Lagrangian to the Hamiltonian approach \cite{Goldstain 1980, Gracia and
Josep, Nikitin-string theory, Rund 1966,Lanczos 1970}.

Fiber bundles provide the mathematical framework for classical mechanics,
field theory, and even quantum mechanics if viewed as a classical field
theory. Parallel transport, covariant differentiation, and gauge symmetry
are very important structures \cite{Pauli 1958} associated with fiber
bundles. When asking: ``What structures are important to physics?'', we
should also ask: ``Why one fiber bundle should be more `physical' than
another?'', ``Why does the `physical' base manifold seem to be a
four-dimensional Lorentzian manifold?'' \cite{Borstnik and Nielsen, van Dam
and Ng, Sachoglu 2001}, and ``How should one construct an action integral
for a given fiber bundle?'' \cite{Kilmister 1967, Carinena 1995, Feynman
1965, Gerjuoy and Rau, Rivas 2001}. Starting with the tangent or cotangent
bundle seems natural because these bundles are related to the notion of a
classical point-like matter. Since we accrue and test our knowledge via
experiments that involve classical apparatus, the physically accessible
fields should be generated by matter and should couple to matter as well.
Therefore, the matter Lagrangian should contain the interaction fields, not
their derivatives, with which classical matter interacts \cite{Dirac 1958}.

In this paper, we discusse the properties of a reparametrization-invariant matter system. 
We use the relativistic particle 
\cite{Rund 1966, Pauli 1958, Feynman 1965, Landau and Lifshitz} to illustrate the main ideas and their generalization to extended objects ($d$-branes). 
We try to find the answer to the question: ``What is the Lagrangian for matter?''
Our matter Lagrangian naturally contains background interaction fields, such as a 1-form field, analogous to the electromagnetic vector potential, and a metric tensor that is usually related to gravity.  We also discuss the guiding principles needed in the construction of
the Lagrangians for the interaction fields.

In Section \ref{Relativistic Particle}, the Lagrangian for a relativistic
particle is given as an example of a reparametrization-invariant action. In
Section \ref{Homogeneous Lagrangians} we argue in favor of first order
homogeneous Lagrangians. Section \ref{New Forces} discuses the physical
implication of such Lagrangians, in particular, the possibility of classical
forces beyond electromagnetism and gravity. Section \ref{Extended Objects}
considers a possible generalization to $D$-dimensional extended objects 
($d$-branes). Section \ref{Field Lagrangians} contains a review of the field
Lagrangians relevant for the interaction fields. Our conclusions and
discussions are given in Section \ref{Conclusions}.

\section{The Matter Lagrangian for The Relativistic Particle}

\label{Relativistic Particle}

From everyday experience, we know that localized particles move with a finite
3D speed. In an extended configuration space (4D space-time), when time is
added as a coordinate ($x^{0}=ct $), particles move with a constant
4-velocity ($v \cdot v=constant$). The 4-velocity is constant due to its
definition $v^{\mu}= dx^{\mu}/d\tau $ that uses the invariance of the proper
time ($\tau $) mathematically defined via a metric tensor $g_{\mu \nu} $ ($d\tau^{2}= g_{\mu \nu}dx^{\mu}dx^{v} $). In this case, the action for a massive
relativistic particle has a nice geometrical meaning: the `distance'
along the particle trajectory \cite{Pauli 1958}: 
\begin{eqnarray}  \label{S1}
S_{1}=\int d\tau L_{1}(x,v)&=&\int d\tau \sqrt{g_{\mu \nu}v^{\mu}v^{\nu}}, \\
\quad \sqrt{g_{\mu \nu}v^{\mu}v^{\nu}}\rightarrow 1&\Rightarrow& S_{1}=\int
d\tau.  \nonumber
\end{eqnarray}
However, for massless particles, such as photons, the length of the
4-velocity is zero ($g_{\mu \nu}v^{\mu}v^{\nu}=0 $). Thus one has to use a
different Lagrangian to avoid problems due to division by zero when
evaluating the final Euler-Lagrange equations. The appropriate `good' action
is \cite{Pauli 1958}: 
\begin{equation}  \label{S2}
S_{2}=\int L_{2}(x,v)d\tau =\int g_{\mu \nu}v^{\mu}v^{\nu}d\tau.
\end{equation}
Notice that the Euler-Lagrange equations obtained from $S_{1} $ and $S_{2} $
are equivalent, and both are equivalent to the geodesic equation as well: 
\begin{eqnarray}  \label{geodesic equations}
\frac{d}{d\tau}\vec{v}=D_{\vec{v}}\vec{v}=v^{\beta}\nabla _{\beta}\vec{v}=0,
\\
\quad v^{\beta}\left(\frac{\partial v^{\alpha}}{\partial x^{\beta}}-\Gamma
_{\beta \gamma}^{\alpha}v^{\gamma}\right) =0.  \nonumber
\end{eqnarray}

In general relativity the Levi-Civita connection $\nabla _{\beta }$, with
Christoffel symbols $\Gamma _{\beta \gamma }^{\alpha }=g^{\alpha \rho
}\left( g_{\beta \gamma ,\rho }-g_{\rho \beta ,\gamma }-g_{\rho \gamma
,\beta }\right) /2$, preserves the length of the vectors 
($\nabla g(\vec{v},\vec{v})=0$) \cite{Pauli 1958}. Therefore, these equivalences are not
surprising because the Lagrangians in ((\ref{S1}) and (\ref{S2})) are
functions of the preserved arc length $g(\vec{v},\vec{v})=\vec{v}^{2}$.
However, the parallel transport for a general connection $\nabla _{\beta }$
does not have to preserve the length of a general vector.

The equivalence between $S_{1} $ and $S_{2} $ is very robust. Since $L_{2} $
is a homogeneous function of order $2$ with respect to $\vec{v} $, the
corresponding Hamiltonian function ($h=v\partial L/\partial v-L $) is
exactly equal to $L $ ($h(x,v)=L(x,v) $). Thus $L_{2} $ is conserved, and so
is the length of $\vec{v} $. Any homogeneous Lagrangian in $\vec{v} $ of
order $n\neq 1 $ is conserved because $h=(n-1)L $. When $dL/d\tau =0 $, then
one can show that the Euler-Lagrange equations for $L $ and 
$\tilde{L}=f\left(L\right) $ are equivalent under certain minor restrictions on $f$.
This is an interesting type of equivalence that applies to homogeneous
Lagrangians ($L(\beta v)=\beta^n L(v)$). It is different from the usual equivalence 
$L\rightarrow \tilde{L}=L+d\Lambda /d\tau $ or the more general equivalence discussed in ref. 
\cite{Hojman and Harleston}. Any solution of the Euler-Lagrange equation for 
$\tilde{L}=L^{\alpha} $ would conserve $L=L_{1} $ since 
$\tilde{h}=(\alpha-1)L^{\alpha} $. All these solutions are solutions of the
Euler-Lagrange equation for $L $ as well; thus $L^{\alpha}\subset L $. In
general, conservation of $L_{1} $ is not guaranteed since $L_{1}\rightarrow
L_{1}+d\Lambda /d\tau $ is also a homogeneous Lagrangian of order one
equivalent to $L_{1} $. This suggests that there may be a choice of $\Lambda 
$, a ``gauge fixing'', so that $L_{1}+d\Lambda /d\tau $ is conserved even if 
$L_{1} $ is not. The above discussion applies to any homogeneous
Lagrangian.

\section{Homogeneous Lagrangians of First Order}
\label{Homogeneous Lagrangians}

Suppose we don't know anything about classical physics, which is mainly
concerned with trajectories of point particles in some space $M $, but we
are told we can derive it from a variational principle if we use the right
action integral $S=\int Ld\tau $. By following the above example we wonder:
``should the smallest `distance' be the guiding principle?'' when
constructing $L$. If yes, ``How should it be defined for other field
theories?'' It seems that a reparametrization-invariant theory can provide
us with a metric-like structure \cite{Rund 1966}, and thus a possible link
between field models and geometric models \cite{Rucker 1977}.

In the example of the relativistic particle, the Lagrangian and the
trajectory parameterization have a geometrical meaning. In general, however,
parameterization of a trajectory is quite arbitrary for any observer. If
there is a smallest time interval that sets a space-time scale, then this
would imply a discrete space-time structure since there may not be any
events in the smallest time interval. The Planck scale is often considered
to be such an essential scale \cite{Magueijo and Smolin}. Leaving aside
recent hints for quantum space-time from loop quantum gravity and other
theories , we ask: ``Should there be any preferred trajectory
parameterization in a smooth 4D space-time?'' and ``Aren't we free to choose
the standard of distance (time, using natural units $c=1 $)?'' If so, then 
\textit{we should have a smooth continuous manifold and our theory should
not depend on the choice of parameterization}.

If we examine the Euler-Lagrange equations carefully: 
\begin{equation}
\frac{d}{d\tau }\left( \frac{\partial L}
{\partial v^{\alpha }}\right) =\frac{\partial L}{\partial x^{\alpha }},  
\label{Euler-Lagrange equations}
\end{equation}
we notice that any homogeneous Lagrangian of order $n$ 
($L(x,\alpha \vec{v})=\alpha ^{n}L(x,\vec{v})$) provides a reparametrization invariance of the
equations under the transformations 
$\tau \rightarrow \tau /\alpha ,\vec{v}\rightarrow \alpha \vec{v}$. 
Next, note that the action $S$ involves an
integration that is a natural structure for orientable manifolds ($M$) with
an $n$-form of the volume. Since a trajectory is a one-dimensional object,
then what we are looking at is an embedding $\phi :\Bbb{R}^{1}\rightarrow M$. 
This means that we push forward the tangential space 
$\phi _{*}:T(\Bbb{R}^{1})=\Bbb{R}^{1}\rightarrow T(M)$, and pull back the cotangent space $\phi
^{*}:T(\Bbb{R}^{1})=\Bbb{R}^{1}\leftarrow T^{*}(M)$. Thus a 1-form $\omega $
on $M$ that is in $T^{*}(M)$ ($\omega =A_{\mu }\left( x\right) dx^{\mu }$)
will be pulled back on $\Bbb{R}^{1}$ ($\phi ^{*}(\omega )$) and there it
should be proportional to the volume form on $\Bbb{R}^{1}$ ($\phi
^{*}(\omega )=A_{\mu }\left( x\right) (dx^{\mu }/d\tau )d\tau \sim d\tau $),
allowing us to integrate $\int \phi ^{*}(\omega )$ : 
\[
\int \phi ^{*}(\omega )=\int Ld\tau =\int A_{\mu }\left( x\right) v^{\mu
}d\tau .
\]

Therefore, by selecting a 1-form $\omega =A_{\mu}\left(x\right) dx^{\mu} $
on $M $ and using $L=A_{\mu}\left(x\right) v^{\mu} $ we are actually solving
for the embedding $\phi:\Bbb{R}^{1}\rightarrow M $ using a chart on $M $
with coordinates $x:M\rightarrow \Bbb{R}^{n}$. The Lagrangian obtained this
way is homogeneous of first order in $v $ with a very simple dynamics. The
corresponding Euler-Lagrange equation is $F_{\nu \mu}v^{\mu}=0 $ where $F $
is a 2-form ($F=dA $); in electrodynamics this is the Faraday's tensor. If
we relax the assumption that $L $ is a pulled back 1-form and assume that it
is just a homogeneous Lagrangian of order one, then we find a
reparametrization-invariant theory that may have an interesting dynamics.

\subsection{Pros and Cons About Homogeneous Lagrangians of First Order}

Although, most of the features listed below are more or less self-evident,
it is important to compile a list of properties of the
homogeneous Lagrangians of first order in the velocity $\vec{v} $.

Some of the good properties of a theory with a first order homogeneous
Lagrangian are:

\begin{itemize}
\item[(1)]  First of all, the action $S=\int L(x,\frac{dx}{d\tau })d\tau $
is a reparametrization invariant.

\item[(2)]  For any Lagrangian $L(x,\frac{dx}{dt})$ one can construct a
reparametrization-invariant Lagrangian by enlarging the space to an extended
space-time \cite{Goldstain 1980}: 
$L(x,\frac{dx}{dt})\rightarrow L(x,\frac{dx}{dt})\frac{dt}{d\tau }$. 
However, it is an open question whether there is a
full equivalence of the corresponding Euler-Lagrange equations.

\item[(3)]  Parameterization-independent path-integral quantization could be
possible since the action $S$ is reparametrization invariant.

\item[(4)]  The reparametrization invariance may help in dealing with
singularities \cite{Kleinert 1989}.

\item[(5)]  It is easily generalized to $D$-dimensional extended objects 
($d$-branes) that is the subject of Section \ref{Extended Objects}.
\end{itemize}

The list of trouble-making properties in a theory with a first order
homogeneous Lagrangian includes:

\begin{itemize}
\item[(1)]  There are constraints among the Euler-Lagrange equations \cite
{Goldstain 1980}, since $\det \left( \frac{\partial ^{2}L}
{\partial v^{\alpha }\partial v^{\beta }}\right) =0$.

\item[(2)]  It follows that the Legendre transformation ($T\left( M\right)
\leftrightarrow T^{*}\left( M\right) $), which exchanges velocity and
momentum coordinates $(x,v)\leftrightarrow (x,p)$, is problematic \cite
{Gracia and Josep}.

\item[(3)]  There is a problem with the canonical quantization approach
since the Hamiltonian function is identically ZERO ($h\equiv 0$) \cite
{Nikitin-string theory}.
\end{itemize}

Constraints among the equations of motion are not an insurmountable problem
since there are procedures for quantizing such theories \cite{Nikitin-string
theory, Dirac 1958a, Teitelboim 1982, Henneaux and Teitelboim, Sundermeyer
1982}. For example, instead of using $h\equiv 0$ one can use some of the
constraint equations available, or a conserved quantity, as Hamiltonian for
the quantization procedure \cite{Nikitin-string theory}. Changing
coordinates $(x,v)\leftrightarrow (x,p)$ seems to be difficult, but it may
be resolved in some special cases by using the assumption that a gauge 
$\Lambda $ has been chosen so that 
$L\rightarrow L+\frac{d\Lambda }{d\tau }=\tilde{L}=const$. We would not 
discuss the above-mentioned quantization
troubles since they are outside of the scope of this paper. A new approach
that resolves $h\equiv 0$ and naturally leads to a Dirac like equation is
under investigation and subject of a forthcoming paper, for some preliminary
details see ref. \cite{VGG Varna 2002}.

\subsection{Canonical Form of the First Order Homogeneous Lagrangians}

By now, we hope that the reader is puzzled, as we are, about the answer to
the following question: ``What is the general mathematical expression for
first order homogeneous functions?'' Below we define what we mean by the 
\textit{canonical form of the first order homogeneous Lagrangian} and why we
prefer such a mathematical expression.

First, note that any symmetric tensor of rank $n $ ($S_{\alpha _{1}\alpha
_{2}...\alpha _{n}}=S_{[\alpha _{1}\alpha _{2}...\alpha _{n}]} $, where 
$[\alpha _{1}\alpha _{2}...\alpha _{n}] $ is an arbitrary permutation of the
indexes) defines a homogeneous function of order $n $ ($S_{n}(\vec{v},..., 
\vec{v})=S_{\alpha _{1}\alpha _{2}...\alpha _{n}}v^{\alpha
_{1}}....v^{\alpha _{n}} $). The symmetric tensor of rank two is denoted by 
$g_{\alpha \beta} $. Using this notation, the canonical form of the first
order homogeneous Lagrangian is defined as: 
\begin{eqnarray}  \label{canonical form}
L\left(\vec{x},\vec{v}\right) &=&\sum_{n=1}^{\infty} \sqrt[n]{S_{n} 
\left(\vec{v},...,\vec{v}\right)}= \\
&=&A_{\alpha}v^{\alpha}+ \sqrt{g_{\alpha\beta}v^{\alpha}v^{\beta}} 
+...\sqrt[m]{S_{m}\left(\vec{v},..., \vec{v}\right)}.  \nonumber
\end{eqnarray}

Whatever is the Lagrangian for matter, it should involve interaction fields
that couple with the velocity $\vec{v} $ to a scalar. Thus we must have 
$L_{matter}\left(\vec{x},\vec{v};Fields ~\Psi \right)$. When the matter action is
combined with the action ($ \int \mathcal{L}[\Psi] dV$) for the interaction fields $\Psi$, 
we obtain a full \textit{background independent theory}. Then the corresponding Euler-Lagrange equations contain ``dynamical derivatives''
on the left hand side and sources on the right hand side: 
\[
\partial _{\gamma}\left(\frac{\delta \mathcal{L}}{\delta
(\partial_{\gamma}\Psi)}\right) =\frac{\delta \mathcal{L}}{\delta \Psi} 
+\frac{\partial L_{matter}}{\partial \Psi}. 
\]

The advantage of the canonical form of the first order homogeneous
Lagrangian (\ref{canonical form}) is that each interaction field, which is
associated with a symmetric tensor, has a unique matter source that is a
monomial in the velocities: 
\begin{equation}
\frac{\partial L}{\partial S_{\alpha _{1}\alpha _{2}...\alpha _{n}}} 
=\frac{1}{n} \left(S_{n}(\vec{v},...,\vec{v})\right) ^{\frac{1-n}{n}}v^{\alpha
_{1}}....v^{\alpha _{n}}.  \label{sources}
\end{equation}

There are many other ways one can write first-order homogeneous functions 
\cite{Rund 1966}. For example, one can consider the following expression 
$L\left( \vec{x},\vec{v}\right) =\left( h_{\alpha \beta }v^{\alpha }v^{\beta
}\right) \left( g_{\alpha \beta }v^{\alpha }v^{\beta }\right) ^{-1/2}$ where 
$h$ and $g$ are seemingly different symmetric tensors. However, each of
these fields ($h$ and $g$) has the same source type ($\sim v^{\alpha
}v^{\beta }$): 
\[
\frac{\partial L}{\partial h_{\alpha \beta }}=\frac{L\left( \vec{x}, \vec{v}
\right) }{h_{\gamma \rho }v^{\gamma }v^{\rho }}v^{\alpha }v^{\beta },\quad 
\frac{\partial L}{\partial g_{\alpha \beta }}=\frac{L\left( \vec{x}, \vec{v}
\right) }{g_{\gamma \rho }v^{\gamma }v^{\rho }}v^{\alpha }v^{\beta }. 
\]
Theories with two metrics have been studied before \cite{Dirac 1979,
Bekenstein 1993}. At this stage, however, we cannot find any good reason why
the same source type should produce different fields. Therefore, we prefer
the canonical form (\ref{canonical form}) for our discussion.

\section{Classical Forces Beyond Electromagnetism and Gravity}
\label{New Forces}

The aim of this paper is to set the stage for diffeomorphism invariant
mechanics of extended objects by close analogy with the relativistic point
particle. However, it is important that we understand the new terms in the
canonical expression of the first order homogenous Lagrangians (\ref
{canonical form}). In this respect this section discusses the implications
of such interaction terms beyond electromagnetism and gravity as given by
the canonical expression of the first order homogeneous Lagrangians (\ref
{canonical form}).

First, we point out that one can circumvent the linear dependence, 
$(\det (\frac{\partial ^{2}L}{\partial v^{\alpha }\partial v^{\beta }})=0)$ due to
the reparametrization symmetry, of the equations of motion derived from 
$L=\sqrt[n]{S_{n}\left( \vec{v},...,\vec{v}\right) }$ by adding an extra set of
equations ($\frac{dL}{d\tau }=0$). This way the equations of motion derived
from $L=\sqrt[n]{S_{n}\left( \vec{v},...,\vec{v}\right) }$ and $\frac{dL}
{d\tau }=0$ are equvalent to the equations of motion derived from 
$L=S_{n}\left( \vec{v},...,\vec{v}\right)$. This is similar to the
discussion at the end of Section \ref{Relativistic Particle}. As noticed before, this
is a specific choice of parametrization such that $v^{\alpha }g_{\alpha
\beta }\left( x\right) v^{\beta }$ is constant. Indeed, if we start with the
re-parametrization invariant Lagrangian 
$L=qA_{\alpha }v^{\alpha }+m\sqrt{g_{\alpha \beta }(x)v^{\alpha }v^{\beta }}$ and define \textbf{proper time} 
$\tau $ such that: $d\tau =\sqrt{g_{\alpha \beta }dx^{\alpha }dx^{\beta }}
\Rightarrow \sqrt{g_{\alpha \beta }v^{\alpha }v^{\beta }}=1$. Then we can
effectively consider $L=qA_{\alpha }v^{\alpha }+(m+\chi )\sqrt{g_{\alpha
\beta }v^{\alpha }v^{\beta }}-\chi $ as our model Lagrangian. Here $\chi $
is a Lagrange multiplier to enforce $\sqrt{g_{\alpha \beta }v^{\alpha
}v^{\beta }}=1$ that breaks the reparametrization invariance explicitly.
Then we can write it as $L=qA_{\alpha }v^{\alpha }+(m+\chi )\frac{g_{\alpha
\beta }v^{\alpha }v^{\beta }}{\sqrt{g_{\alpha \beta }v^{\alpha }v^{\beta }}}
-\chi $ and using $\sqrt{g_{\alpha \beta }v^{\alpha }v^{\beta }}=1$ we get 
$L=qA_{\alpha }v^{\alpha }+(m+\chi )g_{\alpha \beta }v^{\alpha }v^{\beta
}-\chi$. One can fix $\chi $ to be $-m/2$ by requiring that $L=qA_{\alpha
}v^{\alpha }+m\sqrt{g_{\alpha \beta }(x)v^{\alpha }v^{\beta }}$ and 
$L=qA_{\alpha }v^{\alpha }+(m+\chi )g_{\alpha \beta }v^{\alpha }v^{\beta
}-\chi $ give the same Euler-Lagrange equations. This results in the
familiar equivalent Lagrangian: $L=qA_{\alpha }v^{\alpha }+
\frac{m}{2}g_{\alpha \beta }v^{\alpha }v^{\beta }$.

If we focus on a specific  $n^{th}$-term of re-parametrization invariant Lagrangian (\ref
{canonical form}), that is, $L=\left( S_{n}\left( v\right) \right) ^{1/n}$
in the parametrization gauge $S_{n}\left( v\right) =const$ then the
equations of motion are: 
\[
S_{n/\alpha /\beta }(v)\frac{dv^{\beta }}{d\tau }=S_{n,\alpha
}(v)-S_{n/\alpha ,\beta }v^{\beta }.
\]
Here $S_{n,\alpha }$ denotes partial derivative with respect to $x^{\alpha }$
when $S_{n/\alpha }$denotes partial derivative with respect to $v^{\alpha }$.
From this expression it is clear that $n=2$ is a model that results in
velocity independent symmetric tensor $S_{n/\alpha /\beta }(v)$ that can be
associated with the metric tensor. In general $S_{n/\alpha /\beta }(v)$ goes
as $v^{n-2}$ which will result in an interesting behavior for $n>2$: at
velocities that approach zero ($v\rightarrow 0$) the acceleration grows as 
$\frac{1}{v^{n-2}}$.

To further illustrate our point and to gain better understanding of the 
$S_{n}(v)$ terms we assume:

\begin{itemize}
\item  Rotational symmetry, that is: $S_{n}(v)=f(t,r,w,v)$ where $w=dt/d\tau$
and $v=dr/d\tau$,
\item  Static fields, that is: $S_{n}(v)=f(r,w,v)$,
\item  Inertial coordinate system in the sense of Newtonian like space and
time separation, that is: $S_{t...tr...r}=0$ except for $S_{t...t}$ and 
$S_{r...r}$ components:
\end{itemize}
\[
S_{n}(v)=\psi (r)w^{n}+\phi (r)v^{n}.
\]
This way the corresponding equations of motion for $L=S_{n}(v)$ are: 
\[
\frac{dv}{d\tau }=-\frac{v^{2}\phi ^{\prime }(r)}{n\phi (r)}+\frac{1}{v^{n-2}}
\frac{w^{n}\psi ^{\prime }(r)}{(n-1)\phi (r)}, \frac{dw}{d\tau }=
-\frac{wv\psi ^{\prime }(r)}{(n-1)\psi (r)}.
\]

The physics interpretation of such equations of motion is that an observer
cannot study a particle that is in absolute rest with respect to the
observer because this would mean that such particle has an infinite
acceleration. This sounds very similar to the uncertainty principle in
quantum mechanics. Such terms with $n>2$ play important role in the
derivation of the Dirac equation via $v\rightarrow \gamma $ quantization
which will be discussed in the second part of this article (for preliminary
results see \cite{VGG Varna 2002} and \cite{VGG Cincinnati 2003}).

Not being able to observe a particle at rest seems somewhat in contradiction
to our classical physics reality. However, the more appropriate Lagrangian
should take into account that `empty space' has Minkowski geometry: 
\[
L=m\sqrt{\eta _{\alpha \beta }v^{\alpha }v^{\beta }}+\delta \sqrt[n]
{S_{n}\left( \vec{v},...,\vec{v}\right) }.
\]
Here $\eta _{\alpha \beta }=(1,-1,...,-1)$ is the Lorentz invariant metric
tensor. For Lagrangians that contain gravity ($S_{2}(v)$ term) the problem
for special velocity limit $v\rightarrow 0$ does not exist.  In the non-relativistic limit ($v\rightarrow 0$), the present model of pure $S_{n}$ interaction in Minkowski spacetime results in acceleration $\frac{dv}{d\tau }$ that is the same up to $O(v^{2})$ terms for $L=const$ parametrization as well as for $\sqrt{\eta _{\alpha \beta
}v^{\alpha }v^{\beta }}=const$ parametrization. Thus the non-relativistic limit cannot distinguish these two choices of parametrization.

It was mentioned before that for homogeneous Lagrangians of order $\alpha$
we have $H=(\alpha -1)L$ and thus $dL/d\lambda =0$ except for $\alpha =1$
that singles out homogeneous Lagrangians of first order. When working with
re-parametrization invariant Lagrangian, one can chose parametrization so
that $Ld\lambda =d\tau $ or effectively $L(x,v)=const$. This brings the
homogeneous Lagrangians of first order back in the family $dL/d\lambda =0$.
If we don't know the structure of $L$ this seems to be the choice to be made.

It seems, however, that $\sqrt{g_{\alpha \beta }v^{\alpha }v^{\beta }}=const$
is preferred as physically more relevant due to its connection to the lifetime 
of elementary particles. Especially, due to the lack of experimental
evidence that lifetime of charged elementary particle is affected by the
presence of electromagnetic fields. This can be related to the observation
that for any Lagrangian of the form $L=v^{\mu }A_{\mu }(x,v),$ where $x$ is
space-time coordinate and $v$ is a world-line velocity vector (4-vector for
3+1 space-time), one can define a velocity dependent metric $g_{\alpha \beta
}\left( x,v\right) =A_{\alpha /\beta }\left( x,v\right) +A_{\beta /\alpha
}\left( x,v\right) $ where $A_{\beta /\alpha }\left( x,v\right) $ denotes
partial derivative with respect to $v^{\alpha }$ of $A_{\beta }\left(
x,v\right)$. Then one can show that $\frac{d}{d\lambda }\left( v^{\alpha
}g_{\alpha \beta }\left( x,v\right) v^{\beta }\right) =0$ along the
trajectory determined by the Euler-Lagrange equation for $L=v^{\mu }A_{\mu
}(x,v)$. This metric $g_{\alpha \beta }\left( x,v\right) $ does not depend
on the velocity independent electromagnetic vector potential $A_{\mu }(x)$
and thus the length of the vector as calculated with $g_{\alpha \beta
}\left( x,v\right) $ is not affected by the presence of electromagnetic
interaction. For homogeneous Lagrangians of first order, however, one has 
$v^{\alpha }g_{\alpha \beta }\left( x,v\right) v^{\beta }=0$ because $A_{\mu
}(x,v)$ is a homogeneous function of zero degree and thus $v^{\beta }A_{\mu
/\beta }\left( x,v\right) =0$. 

In this respect for homogeneous Lagrangians of first order, it is not clear
if one has to chose the parametrization so that $L=const,$ or 
$\sqrt{g_{\alpha \beta }v^{\alpha }v^{\beta }}=const,$ or $L-A_{\mu }(x)v^{\mu
}=const$. The choice $L-A_{\mu }(x)v^{\mu }=const$ may very well be the
appropriate choice since the weak and the strong forces do have effect on
the lifetime of elementary particles; for example, neutrons are unstable in
free space but stable within the nuclei. In connection to this we note that
the other terms beyond gravity ( $S_{n}$ with $n>2$) are seemingly related
to the internal degrees of freedom of the elementary particles. This becomes
more clear once a non-commutative quantization ( $v\rightarrow \gamma $) is
applied to the re-parametrization invariant Lagrangian. This, however, will
be discussed elsewhere, for some preliminary results see \cite{VGG Varna 2002}
and \cite{VGG Cincinnati 2003}.

To conclude this section, one may naively extrapolate the scale at which
such new forces may be dominant. Considering that electromagnetic forces are
relevant at atomic and molecular scale, when gravity is dominating the solar
system and galactic scales, then one may deduce that terms beyond
gravity may be relevant at galactic and intergalactic scales. Along this
line, a possible determination of the structure of such forces from the
velocity distribution of stars in galaxies is an interesting possibility. In this
respect, such forces can be of relevance to the dark matter and dark energy
cosmology problems. The pathological $dv/d\tau \rightarrow \infty $ when 
$v\rightarrow 0$ behavior of pure $S_{n}$ for $n>2$ interactions could also 
be of relevance to inflation models. Finally, as already mentioned, such terms
are essential in the discussion of the Dirac equation when we consider the
quantization of the homogeneous Lagrangians of first order.

\section{$D$-dimensional Extended Objects}

\label{Extended Objects}

In the previous sections, we have discussed the classical mechanics of a
point-like particle as a problem concerned with the embedding 
$\phi :\Bbb{R}^{1}\rightarrow M$. The map $\phi $ provides the trajectory (the word line)
of the particle in the target space $M$. In this sense, we are dealing with
a $0$-brane that is a one dimensional object. Although time is kept in mind
as an extra dimension, we do not insist on any special structure associate
with a time flow. We think of an extended object as a manifold $D$ with
dimension, denoted also by $D,\dim D=D=d+1$ where $d=0,1,2,...$. In this
sense, we have to solve for $\phi :D\rightarrow M$ such that some action
integral is minimized. From this point of view, we are dealing with
mechanics of a $d$-brane. In other words, how is this $D$-dimensional
extended object submerged in $M$, and what are the relevant interaction
fields? By using the coordinate charts on $M$ ($x:M\rightarrow \Bbb{R}^{m}$), 
we also can think of this as a field theory over the $D$-manifold with a
local fiber $\Bbb{R}^{m}$. Thus the field $\vec{\phi}$ is such that $\phi
^{\alpha }=x\circ \phi :D\rightarrow M\rightarrow \Bbb{R}^{m}$. Following
the relativistic point particle discussion, we consider the space of the 
$D$-forms over the manifold $M$, denoted by $\Lambda ^{D}\left( M\right) $,
that has dimension $\binom{m}{D}=\frac{m!}{D!(m-D)!}$. An element $\Omega $
in $\Lambda ^{D}\left( M\right) $ has the form $\Omega =\Omega _{\alpha
_{1}...\alpha _{D}}dx^{\alpha _{1}}\wedge dx^{\alpha _{2}}\wedge
...dx^{\alpha _{D}}$. We use an arbitrary label $\Gamma $ to index different 
$D$-forms over $M,\Gamma =1,2,...,\binom{m}{D}$; thus $\Omega \rightarrow
\Omega ^{\Gamma }=\Omega _{\alpha _{1}...\alpha _{D}}^{\Gamma }dx^{\alpha
_{1}}\wedge dx^{\alpha _{2}}\wedge ...dx^{\alpha _{D}}$. Next we introduce 
``\textit{generalized velocity vectors}'' with components $\omega ^{\Gamma }$
: 
\begin{eqnarray}
\omega ^{\Gamma } &=&\frac{\Omega ^{\Gamma }}{dz}=\Omega _{\alpha
_{1}...\alpha _{D}}^{\Gamma }\frac{\partial \left( x^{\alpha _{1}}x^{\alpha
_{2}}...x^{\alpha _{D}}\right) }{\partial (z^{1}z^{2}...z^{D})},  \nonumber
\\
\quad dz &=&dz^{1}\wedge dz^{2}\wedge ...\wedge dz^{D}.  \nonumber
\end{eqnarray}
In the above expression, $\frac{\partial \left( x^{\alpha _{1}}x^{\alpha
_{2}}...x^{\alpha _{D}}\right) }{\partial (z^{1}z^{2}...z^{D})}$ represents
the Jacobian of the transformation from coordinates $\{x^{\alpha }\}$ over
the manifold $M$ to coordinates $\{z^{a}\}$ over the $d$-brane. The pull
back of a $D$-form $\Omega ^{\Gamma }$ must be proportional to the volume
form over the $d$-brane: 
\begin{eqnarray}
\phi ^{*}\left( \Omega ^{\Gamma }\right) &=&\omega ^{\Gamma }dz^{1}\wedge
dz^{2}\wedge ...\wedge dz^{D}=  \nonumber \\
&=&\Omega _{\alpha _{1}...\alpha _{D}}^{\Gamma }\frac{\partial \left(
x^{\alpha _{1}}x^{\alpha _{2}}...x^{\alpha _{D}}\right) }{\partial
(z^{1}z^{2}...z^{D})}dz^{1}\wedge dz^{2}\wedge ...\wedge dz^{D}.  \nonumber
\end{eqnarray}
Therefore, it is suitable for integration over the $D$-manifold. Thus the
action for $\phi $ is 
\[
S\left[ \phi \right] =\int_{D}L\left( \vec{\phi},\vec{\omega}\right)
dz=\int_{D}\phi ^{*}\left( \Omega \right) =\int_{D}A_{\Gamma } (\vec{\phi})
\omega ^{\Gamma }dz. 
\]
This is a homogeneous function in $\omega $ and is reparametrization
(diffeomorphism) invariant with respect to the diffeomorphisms of the 
$D$-manifold. If we relax the linearity $L(\vec{\phi},\vec{\omega})=\phi
^{*}\left( \Omega \right) =A_{\Gamma }(\vec{\phi})\omega ^{\Gamma }$ in 
$\vec{\omega}$, then the canonical expression for the homogeneous Lagrangian
is: 
\begin{eqnarray}
L\left( \vec{\phi},\vec{\omega}\right) &=&\sum_{n=1}^{\infty }\sqrt[n]
{S_{n}\left( \vec{\omega},...,\vec{\omega}\right) }=
\label{canonical d-brane L} \\
&=&A_{\Gamma }\omega ^{\Gamma }+\sqrt{g_{\Gamma _{1}\Gamma _{2}}\omega
^{\Gamma _{1}}\omega ^{\Gamma _{2}}}+...\sqrt[m]{S_{m} \left( \vec{\omega}
,...,\vec{\omega}\right) }.  \nonumber
\end{eqnarray}

At this point, there is a strong analogy between the relativistic point
particle and the $d $-brane. However, there is a difference in the number of
components; $\vec{x}, \vec{v}$, and $\vec{\phi}=\vec{x}\circ \phi $ have the
same number of components, but the ``generalized velocity'' $\vec{\omega} $
has $\binom{\dim M}{\dim D} $ components which are Jacobians \cite{Fairlie
and Ueno}.

Some specific examples of $d$-brane theories correspond to the following
familiar Lagrangians:

\begin{itemize}
\item  The Lagrangian for a 0-brane (relativistic point particle in an
electromagnetic field, $\dim D=1$ and $\omega ^{\Gamma }\rightarrow
v^{\alpha }=\frac{dx^{\alpha }}{d\tau }$) is: 
\begin{eqnarray}
L\left( \vec{\phi},\vec{\omega}\right) =A_{\Gamma }\omega ^{\Gamma }
+\sqrt{g_{\Gamma _{1}\Gamma _{2}}\omega ^{\Gamma _{1}}\omega ^{\Gamma _{2}}}
\rightarrow L\left( \vec{x},\vec{v}\right)   \nonumber \\
L\left( \vec{x},\vec{v}\right) =qA_{\alpha }v^{\alpha }+m\sqrt{g_{\alpha
\beta }v^{\alpha }v^{\beta }}.  \nonumber
\end{eqnarray}

\item  The Lagrangian for a 1-brane (strings, $\dim D=2$) \cite
{Nikitin-string theory} is: 
\[
L\left( x^{\alpha },\partial _{i}x^{\beta }\right) =\sqrt{Y^{\alpha \beta
}Y_{\alpha \beta }},
\]
using the notation: 
\begin{eqnarray}
\omega ^{\Gamma }\rightarrow Y^{\alpha \beta }=\frac{\partial (x^{\alpha
},x^{\beta })}{\partial (\tau ,\sigma )}=\det \left( 
\begin{array}{cc}
\partial _{\tau }x^{\alpha } & \partial _{\sigma }x^{\alpha } \\ 
\partial _{\tau }x^{\beta } & \partial _{\sigma }x^{\beta }
\end{array}
\right) =  \nonumber \\
=\partial _{\tau }x^{\alpha }\partial _{\sigma }x^{\beta }-\partial _{\sigma
}x^{\alpha }\partial _{\tau }x^{\beta }.  \nonumber
\end{eqnarray}

\item  The Lagrangian for a $d$-brane has the Dirac-Nambu-Goto term (DNG) 
\cite{Pavsic 2001}: 
\[
L\left( x^{\alpha },\partial _{D}x^{\beta }\right) =\sqrt{Y^{\Gamma
}Y_{\Gamma }}.
\]
\end{itemize}

Notice that most of the Lagrangians above, except for the relativistic
particle, are restricted only to gravity-like interactions. In the case of
the charged relativistic particle, the electromagnetic interaction is very
important. The corresponding interaction term for d-banes is know as
Wess-Zumino term \cite{Bozhilov 2002}.

\section{The Background Fields and Their Lagrangians}

\label{Field Lagrangians}

The uniqueness of the interaction fields and their source types
has been essential for the selection of the matter Lagrangian (\ref
{canonical d-brane L}). The first two terms in the Lagrangian are easily
identified as electromagnetic and gravitational interaction. The other terms
describe new classical forces. It is not yet clear if these new terms are
actually present in nature or not, so we will not engage them actively in the following
discussion, but our aim is to start preparing the stage for such research. 
At this point, we have a theory with background fields since we
don't know the equations for the interaction fields. To complete the theory,
we need to introduce actions for these interaction fields.

One way to write the action integrals for the interaction fields $S_{n}$ in 
(\ref{canonical d-brane L}) follows the case of the $d$-brane discussion.
There, we have been solving for $\phi :D\rightarrow M$ by selecting a
Lagrangian that is more than a pull back of a $D$-form over the manifold $M$. 
In a similar way, we may view $S_{n}$ as an $M$-brane field theory, where 
$S_{n}:M\rightarrow S_{n}M$ and $S_{n}M$ is the fiber of symmetric tensors of
rank $n$ over $M$. This approach, however, cannot terminate itself since new
interaction fields would be generated as in the case of $\phi :D\rightarrow M
$.

Another way assumes that $A_{\Gamma} $ is an $n$-form. Thus we may use the
external algebra structure $\Lambda \left(T^{*}M\right) $ over $M $ to
construct objects proportional to the volume form over $M $. For any $n$
-form $(A)$ objects proportional to the volume form $\Omega _{\rm Vol} $ can be
constructed by using operations in $\Lambda\left(T^{*}M\right)$, such as the external derivative $d$, external multiplication $\wedge$, and the Hodge dual $* $. For example, 
$A\wedge *A $ and $dA\wedge *dA $ are forms proportional to the volume form.

The next important ingredient comes from the symmetry in the matter
equation. That is, if there is a transformation $A\rightarrow A^{\prime} $
that leaves the matter equations unchanged, then there is no way to
distinguish $A $ and $A^{\prime} $. Thus the action for the field $A $
should obey the same symmetry (gauge symmetry).

For example, the matter equation for 4D electromagnetic interaction is 
$d\vec{v}/d\tau =F\cdot \vec{v}$ where $F$ is the 2-form obtained by
differentiation of the 1-form $A$ ($F=dA$), and the gauge symmetry for $A$
is $A\rightarrow A^{\prime }=A+df$. The reasonable terms for a 1-form field
in the field Lagrangian $\mathcal{L}(A)$ are: $A\wedge *A,$ $dA\wedge dA$,
and $dA\wedge *dA$. The first term does not conform with the gauge symmetry 
$A\rightarrow A^{\prime }=A+df$ and the second term $(dA\wedge dA)$ is a
boundary term since $dA\wedge dA=d\left( A\wedge dA\right) $ that gives 
$\int_{M}d\left( A\wedge dA\right) =A\wedge dA$ at the boundary of $M$. This
term is interesting in the quantum Hall effect. Therefore, we are left with
a unique action for electromagnetism: 
\[
S\left[ A\right] =\int_{M}dA\wedge *dA=\int_{M}F\wedge *F. 
\]

For our next example, we look at the terms in the matter equation that
involve gravity. There are two possible choices of matter equation. The
first one is the geodesic equation $d\vec{v}/d\tau =\vec{v}\cdot \Gamma
\cdot \vec{v}$ where $\Gamma $ is considered as a connection 1-form that
transforms in the usual way $\Gamma \rightarrow \Gamma +\partial g$ under
coordinate transformations with the group element $g$. This type of
transformation, however, is not a `good' symmetry since restricting $\Gamma
\rightarrow \Gamma +\Sigma $ to transformations $\Sigma =\partial g,$ such
that $\vec{v}\cdot \Sigma \cdot \vec{v}=0$, would mean to select a subset of
coordinate systems, inertial systems, for which the action $S$ is well
defined and satisfies $S\left[ \Gamma \right] =$ $S\left[ \Gamma +\Sigma
\right] $. Selecting a class of coordinate systems for the description of
a system is not desirable, so we shall not follow this road.

In general, the Euler-Lagrange equations assume a background observer who
defines the coordinate system. For electromagnetism, this is tolerable since
neutral particles are such privileged observers. In gravity, however, there
is no such observer, and the equation for matter should be relational. Such
an equation is the equation of the geodesic deviation: $d^{2}\vec{\xi}/d\tau
^{2}=R\cdot \vec{\xi}$, where $R$ is a Lie algebra $(TM)$ valued
curvature 2-form $R=d\Gamma +[\Gamma ,\Gamma ]$. A general curvature 2-form
is denoted by $F\rightarrow $ $\left( F_{\alpha \beta }\right) _{j}^{i}$.
Here, $\alpha $ and $\beta $ are related to the tangental space $(TM)$ of the base
manifold $M$. The $i$ and $j$ are related to the fiber structure of the
bundle over $M$ where the connection that defines $\left( F_{\alpha \beta }\right)
_{j}^{i}$ is given. Clearly, the Ricci tensor $R$ is a very special
curvature because all of its indices are of $TM$ type. For that reason, it
is possible to contract the fiber degree of freedom with the base manifold
degree of freedom (indices). Thus an action linear in $R$ is possible. In
general, one needs to consider a quadratic action, i.e. trace of $F\wedge *F$
($F_{\alpha \beta j}^{i}\wedge *F_{\alpha \beta i}^{j}$).

Using the symmetries of the Ricci tensor $R $ ($R_{\alpha \beta,\gamma
\rho}=-R_{\beta \alpha,\gamma \rho}=-R_{\alpha \beta,\rho \gamma}=R_{\gamma
\rho,\alpha \beta} $) we have two possible expressions that can be
proportional to the volume form $\Omega$. The first expression is present in
all dimensions and is denoted by $R^{*} $, which means that a Hodge dual
operation has been applied to the second pair of indices ($R_{\alpha
\beta,*(\gamma \rho)} $). The $R^{*} $ action seems to be related to the
Cartan-Einstein action for gravity $S\left[ R\right] =\int R_{\alpha
\beta}\wedge *(dx^{\alpha}\wedge dx^{\beta}) $ \cite{Adak et al 2001}.

The other expression is only possible in a four-dimensional space-time and
involves full anti-symmetrization of $R$ ($R_{\alpha [\beta ,\gamma ]\rho )}$) 
denoted by $R^{\wedge }$. The role and implications of such $R^{\wedge }$
term is not yet clear to us. It would be interesting to study the
renormalizability of a theory with such a term  and if its presence only in
four-dimensional space-time has anything to do with the actual dimension of
the physical space-time. However, a statistical argument \cite{Sachoglu
2001, VGG Kiten 2002} based on geometric and differential structure of
various brane and target spaces seems to be a better explanation for why we
are living in a 4D space-time.

\section{Conclusions and Discussions}

\label{Conclusions}

In summary, we have discussed the structure of the matter Lagrangian ($L $)
for extended objects. Imposing reparametrization invariance of the action $S 
$ naturally leads to a first order homogeneous Lagrangian. In its canonical
form, $L $ contains electromagnetic and gravitational interactions, as well
as interactions that are not clearly identified yet. If one extrapolates
from the strengths of the two known interactions, then one may suggest that
the next terms should be important, if present at all, at big cosmological
scales, such as galactic cluster dynamics. If such forces are not present in nature one needs to understand why is that so. The choice of the canonical
Lagrangian is based on the assumption of one-to-one correspondence between
interaction fields and the type of their sources. If one can show that any
homogeneous function can be written in the canonical form suggested, then
this would be a significant step in our understanding of the fundamental
interactions. Note that an equivalent expression can be considered as well: 
$L=A_{\alpha}(\vec{x},\vec{v})v^{\alpha} $. This expression is simpler, and
is concerned with the structure of the homogeneous functions of order zero 
$A_{\alpha}(\vec{x},\vec{v}) $. If one is going to study the new interaction
fields $S_n, n>2$, then the guiding principles for writing field
Lagrangians, as discussed in the examples of electromagnetism and gravity,
may be useful. It would be interesting to apply the outlined constructions
to general relativity by considering it as a $3 $-brane in a $10 $
dimensional target space ($g_{\alpha \beta}:M\rightarrow S_{2}M $).

\textbf{Acknowledgments.} This work is  partially performed under the auspices of the U. S. Department of Energy by the University of California, Lawrence Livermore National Laboratory under contract No. W-7405-Eng-48. The main research was mostly performed at
Louisiana State University. The author acknowledges helpful discussions with 
Professors R. Haymaker, L. Smolinsky, A. R. P. Rau, P. Kirk, J. Pullin, R. O'Connell, C. Torre, J. Baez, P. Al. Nikolov, E. M. Prodanov, G. Dunne, and L. I. Gould and the financial support from the Department of Physics and Astronomy and the Graduate School at the
Louisiana State University, the U. S. National Science Foundation support
under Grant No. PHY-9970769 and Cooperative Agreement No. EPS-9720652 with matching from the Louisiana Board of Regents Support Fund.

\end{document}